\documentstyle[preprint,revtex]{aps}                  
\def\ltap{\ \raisebox{-.5ex}{\rlap{$\sim$}} \raisebox{.4ex}{$<$}\ }

\newcommand{\be}{\begin{equation}}
\newcommand{\ee}{\end{equation}}
\newcommand{\beq}{\begin{eqnarray}}
\newcommand{\eeq}{\end{eqnarray}}

\begin{document}
\draft
\pagestyle{empty}                                      
\centerline{
                             \hfill   NTUTH--95--03}   
\centerline{\hfill                 April 1995} 
\vfill
\begin{title}
$\mu^+e^- \longleftrightarrow \mu^- e^+$ Transitions via Neutral Scalar Bosons
\end{title}
\vfill
\author{Wei-Shu Hou and Gwo-Guang Wong}
\begin{instit}
Department of Physics, National Taiwan University,
Taipei, Taiwan 10764, R.O.C.
\end{instit}
\receipt{\today}
%
\vfill
\begin{abstract}

With  $\mu\to e\gamma$ decay forbidden by
multiplicative lepton number conservation,
we study muonium--antimuonium transitions induced by
neutral scalar bosons.
Pseudoscalars do not induce conversion
for triplet muonium, while for singlet muonium,
pseudoscalar and scalar contributions
add constructively.
This is in contrast to the usual case of
doubly charged scalar exchange,
where the conversion rate is the same for both
singlet and triplet muonium.
Complementary to muonium conversion studies,
high energy $\mu^+e^- \to \mu^- e^+$
and $e^-e^- \to \mu^- \mu^-$ collisions
could reveal spectacular  resonance peaks
for the cases of neutral and doubly charged scalars, respectively.

\end{abstract}
%
\narrowtext
\pagestyle{plain}


The interest in muonium--antimuonium ($M$--$\bar M$) conversion
dates back to a suggestion by Pontecorvo \cite{Pont},
which pointed out the similarity between the
$M$--$\bar M$ and $K^0$--$\bar K^0$ systems.
Feinberg and Weinberg \cite{FeinWein}
noted  further that
$M$--$\bar M$ conversion is allowed by conservation of
{\it multiplicative} muon number --- muon parity ---
but forbidden by the more traditional additive muon number.
It thus provides a sensitive test of
the underlying conservation law for lepton number(s)
and probes physics beyond the standard model.
One advantage of studying $M$--$\bar M$ conversion
is that, once the effective four-fermion Hamiltonian is given,
everything is readily calculable since it involves just
atomic physics.
The experiment is quite challenging, however,
while on the theoretical front,
it has attracted less attention from model builders
compared to decay modes like $\mu\to e\gamma$
which are in fact forbidden by the multiplicative
law.

The effective Hamiltonian is traditionally taken to be of $(V-A)(V-A)$ form
\begin{equation}
{\cal H}_{M\bar M} = {G_{M\bar M} \over \sqrt{2}}
                                         \bar \mu \gamma_\lambda (1-\gamma_5) e
\,
                                         \bar \mu \gamma^\lambda (1-\gamma_5) e
+ \mbox{H.c.},
\end{equation}
and experimental results are given \cite{PDG} as upper limits on
$R_g \equiv G_{M\bar M}/G_F$, where $G_F$ is the Fermi constant.
The present limit is $R_g < 0.16$ \cite{Matthias}.
This would soon be improved to
$10^{-2}$ level \cite{Jungmann} by
an ongoing experiment \cite{PSI} at PSI,
with the ultimate goal of $10^{-3}$.

Explicit models that lead to effective interactions of eq. (1)
were slow to come by.
In 1982, Halprin \cite{Halprin,Yoshimura} pointed out that
in left-right symmetric (LRS) models with Higgs triplets,
doubly charged scalars $\Delta^{--}$ can mediate
$M$--$\bar M$ transitions at tree level in the $t$-channel (Fig. 1(a)).
The effective interaction, after Fierz rearrangement,
can be put in $(V\pm A)(V\pm A)$ form of eq. (1).
This not only encouraged experimental interests \cite{PDG},
it also stimulated theoretical work 
\cite{H++}.
In particular, Chang and Keung \cite {CK}
give the conditions for a generic model.
These work together gives one the impression that
doubly charged scalar bosons may
be the only credible source for inducing
$M$--$\bar M$ transitions.
However, in a recent model \cite {WH} for radiatively generating
lepton masses from multiple Higgs doublets,
it was pointed out in passing that the
flavor-changing {\it neutral} Higgs bosons responsible for
mass generation could also mediate $M$--$\bar M$ conversion.
A remnant $Z_2$ symmetry serves the
function analogous \cite{CK} to Feinberg-Weinberg's muon parity
that forbids $\mu \to e\gamma$ transitions,
while the effective four-fermion operators
responsible for $M$--$\bar M$ transition
are {\it not} of the form of eq. (1).
In this paper we explore neutral scalar induced
$M$--$\bar M$ oscillations \cite{Derman}
in the general case.
Constraints from $g-2$ and $e^+e^- \to \mu^+\mu^-$
scattering data are studied.
We point out that, complementary to muonium studies,
high energy $\mu^+ e^- \to \mu^- e^+$ and
$e^-e^- \to \mu^-\mu^-$ collisions could clearly distinguish
between (flavor changing) neutral
and doubly charged scalar bosons.


Consider 
neutral scalar and pseudoscalar bosons $H$ and $A$,
with the interaction,
\begin{equation}
-{\cal L}_{Y} = {f_H \over \sqrt{2}}\, \bar \mu e\, H
                     + i {f_A \over \sqrt{2}}\, \bar \mu \gamma_5 e \, A + H.c.
\end{equation}
Imposing a discrete symmetry $P_e$ \cite{CK}
such that the electron as well as  $H$, $A$ fields
are odd while the muon field is even,
processes odd in number of electrons (plus positrons)
like $\mu\to e\gamma$ and $\mu\to ee\bar e$ are forbidden.
Namely, scalar bosons may not possess flavor diagonal
and nondiagonal couplings at the same time.
$P_e$ is nothing but a variation of
the multiplicative muon number of
Feinberg and Weinberg \cite{FeinWein}.
The interaction of eq. (2)
induces (Figs. 1(b) and 1(c)) the effective Hamiltonian
\begin{equation}
{\cal H}_{S,P} = {f_H^2 \over 2m_H^2}\, \bar \mu e\, \bar \mu e
                           - {f_A^2 \over 2m_A^2}\, \bar \mu \gamma_5 e\, \bar
\mu \gamma_5 e,
\end{equation}
at low energy which mediates $M$--$\bar M$ conversion.
The conversion matrix elements for $S^2$ and $P^2$ operators
($S$, $P$ stand for $\bar \mu e$ and
$\bar \mu \gamma_5 e$ densities) are
\begin{eqnarray}
\langle \bar M(F= 0) \vert S^2 \vert  M(F=0) \rangle & = & +{2\over \pi a^3}, \
\ \
\langle \bar M(F= 1)\vert S^2 \vert  M(F=1) \rangle = -{2\over \pi a^3},   \\
\langle \bar M(F= 0)\vert P^2 \vert  M(F=0) \rangle & = & -{4\over \pi a^3}, \
\ \
\langle \bar M(F= 1)\vert P^2 \vert  M(F=1) \rangle = 0,
\end{eqnarray}
where
$F$ is the muonium total angular momentum, while
$a$ is its Bohr radius.
Thus, only scalars induce muonium conversion
in the spin triplet state, while for singlet muonium,
the effect of scalar and pseudoscalar channels
add constructively.
Note that for $(V\pm A)^2$ interactions of eq. (1),
we always get $8G_{M\bar M}/\pi a^3$
for both singlet and triplet muonium \cite{FeinWein}.
One clearly sees that {\it separate} measurements of
singlet vs. triplet $M$--$\bar M$ conversion probabilities
can distinguish between neutral scalar, pseudoscalar
and doubly charged Higgs induced interactions.

In practice, $M$ is formed as a mixture of triplet and singlet states.
It is crucial whether the (anti)muon decays
in the presence of magnetic fields.
Any sizable field strength lifts the
degeneracy of $M$--$\bar M$
for $F = 1$, $m_F = \pm 1$ states,
hence effectively ``quenches" \cite{FeinWein}
the $M$--$\bar M$ conversion.
This is normally the case under realistic conditions,
but experiments correct for this and report $G_{M\bar M}$
(or $R_g$) {\it for zero $B$ field}.
It is important to note, however,
that in so doing, one inadvertantly ignores the
possible differences in the neutral (pseudo)scalar case.
Let us take the example of the ongoing PSI experiment \cite{PSI}.
Muonium  is formed and stays in the
presence of 1kG magnetic field.
In this case,
muonium states are populated as
32\%, 35\%, 18\% and 15\%, respectively, for
$(F,\ m_F) = (0,\ 0),\ (1, +1),\ (1,\ 0),\ (1,\ -1)$.
Only the $m_F = 0$ modes are active for
muonium conversion, hence the effective triplet probability
comes only from $\vert c_{1,0}\vert^2 = 18\%$, down from $68\%$.
For $(V\pm A)^2$ interactions, one simply corrects for
a factor of 1/2 reduction.
For our case of neutral scalar induced interactions,
the experimental limit on $G_{M\bar M}$ relates to
scalar couplings as
\begin{equation}
{G_{M\bar M}^{\rm expt.} \over \sqrt{2}}
= {1\over 8} \sqrt{\vert c_{0,0}\vert^2
                                  \left({f_H^2 \over m_H^2} + 2\, {f_A^2 \over
m_A^2}\right)^2
                            +  \vert c_{1,0}\vert^2 \left(-{f_H^2 \over
m_H^2}\right)^2}.
\end{equation}
Several cases are of interest:
(a) $f_A = 0$;
(b) the ``$U(1)$ limit" of $m_A = m_H$
($H$ and $A$ form a complex neutral scalar),
with  $f_A = f_H$;
(c) $f_H = 0$ (pseudoscalar only).
For case (a),
the result is rather similar to eq. (1).
For case (b), constructive interference
strongly enhances the effect in singlet channel.
For case (c),
only the singlet (0, 0) part is active.
If the PSI experiment will attain \cite{Jungmann} the limit of
$R_g < 10^{-2}$ without observing $M$--$\bar M$ conversion,
eq. (6) would imply the bounds
\begin{equation}
f^2/m^2 \ \raisebox{-.5ex}{\rlap{$\sim$}} \raisebox{.4ex}{$<$}\
            (0.9,\ 0.4,\ 0.6) \times 10^{-6}\ \mbox{\rm GeV}^{-2},
\end{equation}
respectively, for the three cases,
where $f/m$ stand for $f_H/m_H$ except for case (c).

Some other constraints on ${\cal H}_{S,P}$,
such as the anomalous magnetic moments of the
electron and muon, should be considered.
Defining $a \equiv (g - 2)/2$, we find that
\begin{equation}
\delta a_e \simeq - {f^2 \over 16\pi^2}\, m_e
                         \left({m_e\over 3m^2} \mp {3m_\mu\over 2m^2}
                                     \mp {m_\mu\over m^2} \ln {m_e^2\over m^2}
                                           \right),
\end{equation}
where $\mp$ is for $H$ or $A$ contribution, respectively,
while for $a_\mu$ one interchanges $e\longleftrightarrow \mu$.
Comparing experimental measurements \cite{PDG}
with QED prediction,
we find $\delta a_e^{\rm expt.} = (146 \pm 46)\times 10^{-12}$
and $\delta a_\mu^{\rm expt.} = (27 \pm 69)\times 10^{-10}$.
The effective bound from $\delta a_e^{\rm expt.}$
on $f^2/m^2$ is of order $G_F$,
except for the $U(1)$ limit case.
In the latter case, cancellations between $H$ and $A$
lead to a much weaker limit.
However, for muon $g-2$
the leading term (proportional to $m_\mu^2$)
comes from the first term of eq. (8) which
does not suffer from $H$--$A$ cancellation.
Hence, it gives a bound of order $10\ G_F$
for all cases.
In any rate, these limits are considerably weaker than eq. (7).

An interesting constraint comes from high energy
$e^+e^-\to \mu^+\mu^-$ scattering cross sections,
which probe the interference effects between
the contact terms of eq. (2)
(Fig. 1(b) in $t$-channel) and standard diagrams.
For 
case (b),
the effective contact interaction
can be put in standard form \cite{Buskulic}
for compositeness search,
\begin{eqnarray}
{\cal H}_{ee\mu\mu} & = & {f^2 \over 2m^2}\,
(\bar \mu e\, \bar \mu e -  \bar \mu \gamma_5 e\, \bar \mu \gamma_5 e)
\nonumber \\
& = & {g^2 \over 2 \Lambda^2}
\left\{\bar e\gamma_\alpha R e\, \bar \mu\gamma^\alpha L \mu
     +  \bar e\gamma_\alpha L e\, \bar \mu\gamma^\alpha R \mu \right\},
\end{eqnarray}
where $\Lambda \equiv \Lambda^+_{LR}$.
Setting $g^2/(4\pi) = 1$,
the combined limit gives
$\Lambda(ee\mu\mu) > 2.6$ TeV \cite{Buskulic},
which translates to
$f^2/m^2 < 1.9 \times 10^{-6}\ \mbox{\rm GeV}^{-2}$.
This can be converted to a limit on $M$--$\bar M$ conversion
by assuming eq. (6),
\begin{equation}
G_{M\bar M} < 0.06\ G_F,
\end{equation}
which is better than present \cite{Matthias}
$M$--$\bar M$ conversion bound of $R_g < 0.16$,
but somewhat weaker than the $10^{-2}$
bound expected soon at PSI \cite{Jungmann}.


In the model of ref. \cite{WH},
scalar interactions of the type of eq. (2)
were used to generate charged lepton masses
iteratively order by order,
via effective one loop diagrams with
lepton seed masses from one generation higher.
To be as general as possible, we are not concerned with
the generation of $m_\mu$ from $m_\tau$ here.
However, in analogy to the softly broken $Z_8$ symmetry
of ref. \cite{WH},
some discrete symmetry can be invoked to
forbid electron mass at tree level
but allow it to be generated by $m_\mu$
via one loop diagrams as shown in Fig. 2.
Since $m_{H,A} \gg m_\mu$, we have
\begin{equation}
   {m_e\over m_\mu}  \cong  {f^2\over 32\pi^2} \log{m_H^2\over m_A^2}.
\end{equation}
Note that $f_H = f_A = f$ is necessary for divergence cancellation,
hence in the $U(1)$ limit \cite{WH} of $m_A = m_H$
the mass generation mechanism is ineffective.
We see that, because the factor of $1/32\pi^2 \sim 1/300$
is already of order $m_e/m_\mu$,
if $m_A \neq m_H$ but are of similar order of magnitude,
in general we would have $f \sim 1$.
This looks attractive for scalar masses
far above the weak scale since one could
have large Yukawa couplings but
at the same time evade the bound of eq. (7).
However,
in the more ambitious model of ref. \cite{WH},
radiative mass generation mechanism is pinned to the weak scale,
namely, Higgs boson masses cannot be far above TeV scale
for sake of naturalness.
In this case, although eq. (11) still looks attractive and is a
simplified version of the more detailed results of ref. \cite{WH},
with $f \sim 1$ and $m_H,\ m_A \ltap$ TeV,
the bound of eq. (7) cannot be satisfied.
We thus conclude that the bound of eq. (7),
expected soon from PSI, will rule out the possibility
of radiatively generating $m_e$ {\it solely} from $m_\mu$
via one loop diagrams,
if the lepton number changing
neutral scalar bosons are of weak scale mass.
A model where $m_e$ dominantly comes
from $m_\tau$ at one loop level,
with a minor contribution from $m_\mu$,
would be presented elsewhere \cite{HW}.


If $M$--$\bar M$ conversion is observed,
one would certainly have to make separate measurements
in singlet vs. triplet states
to distiniguish between the possible sources.
Complementary to this, one could explore signals at high energies.
It was pointed out already by Glashow \cite{Glashow} the
connection between
the studies of $e^-e^- \to \mu^- \mu^-$ collisions
and $M$--$\bar M$ conversion.
Indeed, shortly after the first $M$--$\bar M$ experiment \cite{Amato},
studies of $e^- e^-$ collisions at SLAC improved
the limit on $G_{M\bar M}$ by a factor of 10 \cite{Barber}.
Although such efforts have not been repeated,
it has been stressed recently by Frampton \cite{Frampton}
in the context of dilepton gauge bosons \cite{dilepton}.
It is clear that if $\Delta^{--}$ exists,
it would appear as a resonance peak
in energetic $e^-e^- \to \mu^-\mu^-$ collisions.

In contrast, it has rarely been mentioned \cite{Yoshimura}
that $\mu^+ e^- \to \mu^- e^+$ collisions may also be of great interest.
Even for $\Delta^{--}$ bosons, the cross section can be
sizable for $\sqrt{s} \sim m_\Delta$.
However, if
neutral scalars that mediate $M$--$\bar M$ conversion
exist, {\it and} the masses are of order TeV or below,
one would have
{\it spectacular $s$-channel resonances in
$\mu^\pm e^\mp$ collisions!}
Even the non-observation of $M$--$\bar M$ conversion
does not preclude this possibility.
Let us assume that PSI would not observe $M$--$\bar M$ conversion
at $10^{-2}$ level, hence $f^2/m^2$ is bound by eq.  (7).
Assuming just a single scalar boson $H$ (case (a))
that saturates such a bound,
and that $H\to \mu^\pm e^\mp$ only,
we plot in Fig. 3 the cross section
$\sigma(\mu^+ e^- \to \mu^- e^+)$ vs.
$\sqrt{s}$ for $m_H = $ 0.25, 0.5, 1 and 2 TeV.
The result for $\Delta^{--}$
constrained by $G_{M\bar M} \ltap 10^{-2}$
is also shown in Fig. 3 as dashed lines
for similar masses.
Note that for $f = 0.1 - 2$, which is the plausible range
for Yukawa couplings advocated in ref. \cite{WH,HW},
eq. (7) implies that the lower bound for $m_H$ ranges between
$100$ GeV -- 2 TeV.
For $e^-e^- \to \mu^- \mu^-$ collisions, the curves are
rather similar, with the role of $H$ and $\Delta^{--}$
interchanged.
It is clear that $\mu^+ e^-$ or $e^- e^-$ colliders in the
few hundred GeV to TeV range have the potential of observing
huge cross sections, and could clearly distinguish between
$H$ and $\Delta^{--}$.

The development of $\mu^+\mu^-$ colliders
have received some attention recently \cite{mumu}.
Perhaps one could also consider the $\mu^\pm e^\mp$ collider option,
especially if one could utilize existing facilities.
As muons are collected via $\pi \to \mu$ decay,
existing accelerator complexes that have both
electron and proton facilities,
such as CERN or HERA, are preferred.
Since $\mu^+$ is easier to collect and cool,
while $e^-$ requires no special effort,
$\mu^+ e^-$ collisions should be easier to perform.
For example, take $E_e$ to be the LEP II beam energy of 90 GeV,
if intense 200 GeV -- 7 TeV $\mu^+$ beams could be produced,
one could attain $\sqrt{s} \simeq$ 190 GeV -- 1.1 TeV.
Compared with problems like $\mu$ decay before collision
for $\mu^+\mu^-$ colliders \cite{mumu},
$\mu^- e^+$ events in $\mu^+ e^-$ collisions
have practically no background.
Future linear colliders should be able to span an even
wider energy range, perhaps performing
$e^- e^-$, $\mu^\pm e^\mp$, $\mu^+\mu^-$ as well as
$e^- e^+$ collisions.


Some discussion is in order.
Neutral scalars with flavor changing couplings
may appear to be exotic \cite{GW}.
However, with multiplicative lepton number,
one evades the bounds from $\mu\to e\gamma$
decay and the like.
In this light, we note that any model with
more than one Higgs doublet in general would
give rise to flavor changing neutral scalars.
Second,
the couplings of eq. (2) demand that $H$ and $A$
carry weak isospin, hence they must have charged partners.
These charged scalars can induce the so-called
``wrong neutrino" decay $\mu^- \to e^-\nu_e\bar \nu_\mu$ \cite{PDG}.
Third,
the conversion matrix elements for
$(S\pm P)^2$ part of eq. (3) can be related to eq. (1), but the
$(S\pm P)(S\mp P)$ parts are related to
$(V\pm A)(V\mp A)$ operators, which were considered by
Fujii {\it et al.} \cite {Fujii} in the context of
dilepton gauge bosons.
In general, $M$--$\bar M$ conversion may have
four different kind of sources:
doubly charged scalar or vector bosons in $t$-channel,
or neutral scalar or vector bosons in $s$- or $t$-channel.
Dilepton gauge boson models are therefore of the second type.
Neutral vector bosons would come from
horizontal gauge symmetries,
but models are somewhat difficult to
construct \cite{HS}.
Detailed measurements of singlet vs. triplet $M$--$\bar M$ conversion,
as well as high energy $\mu^\pm e^\mp \to \mu^\mp e^\pm$
and $e^- e^- \to \mu^-\mu^-$ collisions should be
able to indentify the actual agent for these lepton number
violating interactions.
Fourth,
in supersymmetric theories containing
$R$-parity violating terms \cite{HM},
$s$-channel $\tilde \nu_\tau$
($\tau$ sneutrinos, a kind of neutral scalar) exchange
could also induce $M$--$\bar M$ conversion,
resulting in $(S-P)(S+P)$ operators.

Let us summarize the novel features of this work.
We point out that neutral (pseudo)scalars
may well induce muonium--antimuonium transitions,
something that has been largely neglected in the literature.
All one needs is to invoke multiplicative lepton number
rather than adhering to the traditional but more restrictive
additive lepton number conservation.
In this way, stringent limits from $\mu\to e\gamma$ decay,
etc., are evaded.
The induced operators differ from
the usual $(V-A)(V-A)$ form,
and care has to be taken when one interprets
experimental limits.
In particular, measuring $M$--$\bar M$ conversion
strength in both singlet and triplet muonia can distinguish
between different interactions.
A limit of $G_{M\bar M} < 10^{-2}\ G_F$,
expected soon at PSI,
would rule out the possibility of
radiatively generating $m_e$ solely from $m_\mu$
at one loop order via neutral scalar bosons
with weak scale mass.
Complementary to $M\bar M$ studies,
high energy $\mu^+ e^- \to \mu^- e^+$ collisions
may reveal resonance peaks for flavor changing neutral scalars,
while the more widely known doubly charged scalar
would appear as resonances in $e^- e^- \to \mu^-\mu^-$ collisions.

\acknowledgments
We thank D. Chang and R. N. Mohapatra  for discussions,
and K. Jungmann for numerous communications.
The work of WSH is supported by grant NSC 84-2112-M-002-011,
and GGW by NSC 84-2811-M-002-035
of the Republic of China.

\vskip -1cm
\figure{Diagrams for $\mu^+ e^- \to \mu^- e^+$ transitions via
            (a) doubly charged scalar $\Delta^{--}$;
           and (b), (c) neutral (pseudo)scalars $H$, $A$.}
\vskip -1cm
\figure{One loop diagram for $m_e$ generation.}
\vskip -1cm
\figure{$\sigma(\mu^+ e^- \rightarrow \mu^- e^+)$ {\it vs.}
        $\sqrt{s}$ for $m_H =$ 0.25, 0.5, 1, 2 TeV.
        Only $H\to \mu^\pm e^\mp$ is taken into account for $\Gamma_H$,
        with Yukawa couplings saturating
        $f_H^2/m_H^2 \ltap 0.9\times 10^{-6}$ GeV$^{-2}$.
        Analogous bounds for the case of $\Delta^{--}$ is shown as dashed
lines.}

\eject
\end{document}